\documentclass{article}
\usepackage[preprint]{spconf}
\usepackage{amsmath,graphicx}

\usepackage{tikz}
\usetikzlibrary{shapes, arrows, positioning, fit, calc, math}
\usepackage{pgfplots}
\pgfplotsset{filter discard warning=false,compat=newest,/pgf/number format/.cd,1000 sep={}}
\pgfplotsset{every error bar/.style={line width=1pt}}
\usepgfplotslibrary{colormaps}
\usepackage[separate-uncertainty=true, multi-part-units=single,bracket-numbers = false,bracket-unit-denominator = false, exponent-product = \cdot, round-mode=places]{siunitx}
\usepackage[version=4]{mhchem}

\newcommand*\circled[1]{\tikz[baseline=(char.base)]{\node[shape=circle,draw,inner sep=0.1pt] (char) {#1};}}

\makeatletter
\newcommand\ifexpstrequal[2]{%
  \ifnum\pdfstrcmp{#1}{#2}=0
    \expandafter\@firstoftwo
  \else
    \expandafter\@secondoftwo
  \fi
}
\makeatother

\usepackage{xparse}
\ExplSyntaxOn
\msg_new:nnn
  { switchcase }
  { no-match }
  { There~is~no~entry~`#1'~in~the~switch~statement! }
\NewDocumentCommand \switchcase { m m }
  {
    \str_case:nnF { #1 } { #2 } { \msg_error:nnn { switchcase } { no-match } { #1 } }
  }
\ExplSyntaxOff

\def\mass(#1){%
  \ifexpstrequal{#1}{O}{2.65682E-26}{} % [kg]
  \ifexpstrequal{#1}{O2}{5.31352E-26}{} % [kg]
  \ifexpstrequal{#1}{CO}{4.6513E-26}{} % [kg]
  \ifexpstrequal{#1}{CO2}{7.3081E-26}{} % [kg]
  \ifexpstrequal{#1}{C}{1.99447E-26}{} % [kg]
  }
\def\surfvibtemp(#1){%
  \ifexpstrequal{#1}{O}{213.82}{} % [K]
  \ifexpstrequal{#1}{O2}{26.62}{} % [K]
  \ifexpstrequal{#1}{CO}{67.40}{} % [K]
  \ifexpstrequal{#1}{CO2}{93.85}{} % [K]
  \ifexpstrequal{#1}{C}{377.33}{} % [K]
  }
\def\ActEnergyone(#1){%
  \ifexpstrequal{#1}{O}{4276.38}{} % [K]
  \ifexpstrequal{#1}{O2}{532.49}{} % [K]
  \ifexpstrequal{#1}{O2diss}{2632.26}{} % [K]
  \ifexpstrequal{#1}{CO}{1348.09}{} % [K]
  \ifexpstrequal{#1}{COdiss}{34132.02}{} % [K]
  \ifexpstrequal{#1}{CO2}{1877.08}{} % [K]
  \ifexpstrequal{#1}{CO2diss}{22015.57}{} % [K]
  \ifexpstrequal{#1}{C}{7546.55}{} % [K]
  }
\def\ActEnergytwo(#1){%
  \ifexpstrequal{#1}{O}{3310.40}{} % [K]
  \ifexpstrequal{#1}{O2}{408.53}{} % [K]
  \ifexpstrequal{#1}{O2diss}{8378.67}{} % [K]
  \ifexpstrequal{#1}{CO}{1034.16}{} % [K]
  \ifexpstrequal{#1}{COdiss}{34132.02}{} % [K]
  \ifexpstrequal{#1}{CO2}{1455.60}{} % [K]
  \ifexpstrequal{#1}{CO2diss}{22015.57}{} % [K]
  \ifexpstrequal{#1}{C}{5843.80}{} % [K]
  }
\def\const(#1){%
  \ifexpstrequal{#1}{pi}{3.14159265}{} % [-]
  \ifexpstrequal{#1}{boltzmann}{1.381e-23}{} % [J/K]
  \ifexpstrequal{#1}{planck}{6.626e-34}{} % [J*s]
  \ifexpstrequal{#1}{avogadro}{6.022e23}{} % [1/mol]
  }

\tikzstyle{atop}      = [circle, node distance=40pt, draw, fill=black, text width=10pt, text centered, font=\scriptsize, text=black!20, minimum size=10pt, inner sep=0pt]
\tikzstyle{hollow}    = [circle, node distance=40pt, fill=white, draw=white, text width=10pt, text centered, text=black, font=\scriptsize, minimum size=10pt, inner sep=0pt]
\tikzstyle{hollowhcp} = [circle, node distance=40pt, fill=black!20, draw=black!20, text width=10pt, text centered, text=black, font=\scriptsize, minimum size=10pt, inner sep=0pt]
\tikzstyle{bridge}    = [circle, node distance=40pt, fill=white, draw=white, text width=10pt, text centered, text=black, font=\scriptsize, minimum size=5pt, inner sep=0pt]
\tikzstyle{adsorbate} = [circle, draw, fill=red!40, text centered, minimum size=15pt, text width=10pt, font=\scriptsize, inner sep=0pt]
\tikzstyle{adsorbate2}= [circle, draw=red!60, fill=red!60, text centered, minimum size=10pt, text width=10pt, font=\scriptsize, inner sep=0pt]

\tikzstyle{block} = [rectangle, node distance=2cm, draw, fill=black!10, text width=5em, text centered, rounded corners, minimum height=4em]
\tikzstyle{newblock} = [rectangle, node distance=2cm, draw, fill=red!10, text width=5em, text centered, rounded corners, minimum height=4em]
\tikzstyle{blockinner} = [rectangle, node distance=3cm, draw, text width=5em, text centered, rounded corners, minimum height=4em]
\tikzstyle{cloudinner} = [draw, rectangle, dashed]
\tikzstyle{line} = [draw, -latex']

\newcounter{tblEqCounter} %create a counter

\newcounter{owncounter}

\title{Modelling and Simulation of Heterogeneous Reactions with Statistical Particle Methods}
\name{W. Reschke, M. Pfeiffer and S. Fasoulas}
\address{Institute of Space Systems\\
         University of Stuttgart, Pfaffenwaldring 29, 70569 Stuttgart, Germany}

\begin{document}
\maketitle
\begin{abstract}
  Estimating the heat loads on re-entry vehicles is a crucial part of preparing for atmospheric re-entry manoeuvres.
  Re-entry flows at high altitudes are in the rarefied regime and are governed by high enthalpies and thermodynamic non-equilibrium.
  Additionally, catalytic gas-surface reactions change the gas flow composition and can have a major influence on the heat transfer.
  Our goal is to estimate the heat loads without a priori fitting of simulation parameters to experiments.
  We use the tool PICLas for simulations of such rarefied gas flows.
  It combines different particle methods, including the Direct Simulation Monte Carlo method, for modelling of gases.
  Recently it has been extended to include different catalysis models to treat reactions on surfaces.
  We evaluate a kinetic Monte Carlo approach to model catalytic gas-surface interactions in combination with flow simulations using particle methods.
  Here, the adsorbate distribution is modelled by reproducing a surface system using a kinetic Monte Carlo approach and estimating the necessary parameters using model assumptions.
  This catalytic model is compared to a simple recombination model.
  We present simulations that show the capability of the implemented models for a \ce{SiO2} surface in an Oxygen flow.
  Furthermore, simulation results are compared to heat fluxes and recombination coefficient obtained from the respective experiment.
  The results show that simulations using the kinetic Monte Carlo approach match the experimentally obtained values.
  Thus, the approach can be used to estimate the reactivity of oxygen flows over \ce{SiO2} surfaces.
\end{abstract}

\begin{keywords}
Plasma flow, PICLas, DSMC, Catalysis, Monte Carlo, Surface modelling, KMCS
\end{keywords}

\section{Introduction}
In plasma flow applications, such as atmospheric entries, high forces and heat fluxes are exerted on the thermal protection system (TPS) of entry vehicles.
Since the flow is very reactive, a large amount of the heat flux is resulting from reactions of the gas with the TPS surface.
Therefore, heterogeneous reactions have to be considered for design purposes of TPS.
These reactions involve different interaction types~\cite{kolasinski2012surface} that have to be considered.
%In this paper, reactions of adsorbed particles on the surface, such as Langmuir-Hinshelwood reaction, molecular desorption, diffusion, and dissociation of an adsorbed molecule, will be referred to as desorption mechanisms.
%Accordingly, reactions of gas-phase particles with the surface or adsorbed particles, such as dissociative adsorption, molecular adsorption or Eley-Rideal reaction, will be referred to as adsorption mechanisms.

Flow predictions for design purposes of TPS are made either on the basis of experiments or numerical simulations.
Experiments reflect the real physics, however, they do not provide a complete insight into all phenomena for a full scale problem.
This results from many simultaneously occurring interactions and their mutual dependencies, which can not be measured with sufficient accuracy.
In addition, the same conditions of the full (real) scale problem can rarely be reproduced by experiments due to technical restrictions.
In contrast, numerical simulations typically exhibit improved result and phenomena evaluation.
However, they can only reflect the physical problem as accurately as the used models.
Consequently, simulations may yield inaccurate results or predict unrealistic behaviour.

Catalytic flows near thermal and chemical non equilibrium have been investigated by few research groups~\cite{fertig2005modellierung, padilla2007assessment, laux_diss, molchanova2018surface}.
The focus was mainly on describing surface interactions by modelling rate equations and fitting parameters to experiments.
Our goal is to be able to simulate a broader range of flow cases and surfaces without a priori parameter fitting to experiments.
Here, we focus on two catalytic models.
The first is a simple recombination model~\cite{fasoulas_diss}
The second is a microscopically driven kinetic surface description.

In this work, we use PICLas~\cite{piclas} to simulate a catalytic plasma flow~\cite{massuti_jtht2017} with enabled catalytic models.
PICLas enables numerical simulations of rarefied plasma flows using particle methods such as the Direct Simulation Monte Carlo (DSMC)~\cite{bird1994molecular} and Particle In Cell (PIC) method~\cite{pic_dsmc_ueberblick}.
The modular framework of PICLas allows for the implementation of other particle methods, besides PIC and DSMC.
Here, we focus on the DMSC method when simulating the gas flow.
The DSMC method approximates the collision term of the Boltzmann equation considering binary collisions.
%In each computation cell all particles are paired before collisions are computed.
%Pairing is performed using the nearest neighbour approach with an octree sub-cell splitting~\cite{pfeiffer2013grid}.
%Internal energy exchange (relaxation) and chemical reactions are treated statistically through phenomenological models~\cite{pfeiffer2016direct}.
Surface interactions are mos commonly treated with an extended Maxwell scattering model~\cite{maxwell1878stresses}.
The surface modelling approach in PICLas has been extended with catalytic models to treat heterogeneous reactions~\cite{reschke_rgd31}.

In the first chapter we describe the approach for the modelling of the gas flow and the Direct Simulation Monte Carlo method.
In the second chapter we illustrate the catalytic models implemented in PICLas.
In the third we give an overview of PICLas.
Subsequently, the simulation setup is described and simulation results are shown and discussed.

\section{Direct Simulation Monte Carlo method}
The solution of the Boltzmann equation, especially the collision integral, can be treated by the Direct Simulation Monte Carlo method.
The Boltzmann equation describes the microscopic gas state and the changes of the microscopic distribution.
\begin{equation}
  \frac{\partial f}{\partial t}+\vec{v}\cdot\nabla_{\vec{x}} \, f=
  \frac{\partial f}{\partial t}\bigg|_{Coll}
  \label{eq:boltzmann}
\end{equation}
Here, $f(\vec{x},\vec{v},t)$ is the particle distribution function dependent on the particles' position $\vec{x}$, velocity $\vec{v}$,
time $t$, and $\frac{\partial f}{\partial t}|_{Coll}$ representing the Boltzmann collision integral accounting for binary inter-particle collisions.
This equation is solved by directly simulating particles and their interactions with one another.
Herein, the particle distribution function of the gas system is statistically approximated with $N_{\mathrm{p}}$ representative discrete particles distributed in space
The particle motion and inter particle collisions are separated.
To reduce computational effort not all real molecules $N_{\mathrm{real}}$ are simulated but the simulation particles $N_{\mathrm{sim}}$ are weighted with $W_{\mathrm{p}} = N_{\mathrm{real}} / N_{\mathrm{sim}}$.
The simulation time step is chosen to resolve the collision frequency of the simulated gases~\cite{bird1994molecular}.
Within DSMC, particle collisions are treated statistically and the appropriate probabilities are calculated with phenomenological models~\cite{borgnakke1975statistical}.
Each computational cell must contain at least four simulation particles for representative collision treatment and sampling purposes.
Before the collision treatment the particles are paired.
This is done locally within each computational cell using either a statistical or a nearest neighbour approach.
For statistical particle pairing the mesh cell sizes are determined by the mean free path $\lambda$ to ensure physically meaningful particle separation distances.
The nearest neighbour pairing does not depend on cell sizes, but the simulated particle mean separation distance has to be lower than $\lambda$.
Additionally, an octree cell-subdivision is implemented~\cite{pfeiffer2013grid} to be grid independent for statistical pairing and reduce computational cost for the nearest neighbour approach.
Collision processes lead to momentum and energy exchange (relaxation) as well as chemical reactions.
These include dissociation, recombination and exchange reactions, that are treated with an Arrhenius based approach~\cite{bird1994molecular}.
Finally, macroscopic properties that are of interest such as pressure and temperature are calculated as moments from the sampled resulting particle statistics.

\section{Surface reaction models}
The first of the considered models is a simple recombination approach.
Gas particles that collide with surfaces have a chance to recombine, e.g. $\ce{O}^{(\mathrm{gas})} + \ce{O}^{(\mathrm{surf})} \rightarrow \ce{O2}^{(\mathrm{gas})}$.

The second approach uses a kinetic Monte Carlo surface (KMCS) model for gas-surface reactions.
This results in a more complex reaction treatment that also considers dissociation and exchange reactions.
In this work, only a short overview of the basic functionality is given. A more detailed description will be available in a following publication.

\subsection{Simple Recombination Model}
The first model is an extension of the Maxwell scattering on surfaces and has a straightforward implementation within a DSMC framework.
Here, an additional coefficient $\gamma$ is added, defining the ratio of recombining $N_{\mathrm{rec}}$ to incident atoms $N_{\mathrm{i}}$~\cite{fasoulas_diss}. 
For $ \gamma = 0 $ the surface is defined to be non-catalytic and fully catalytic for $ \gamma = 1 $.
The corresponding probabilities are chosen for each surface collision with
\begin{eqnarray}
  \label{eq:recomb_adsorb}
  P_{\mathrm{ads},i} &=& \gamma_{i} - P_{\mathrm{rec},i}, \\
  \label{eq:recomb_recomb}
  P_{\mathrm{rec},i} &=& \gamma_{i} \, \left(1-exp(-N_{\mathrm{s}})\right),
\end{eqnarray}
such that they reduce statistical noise~\cite{laux_diss}.
Here, $N_{\mathrm{s}}$ is the number of particles that are adsorbed on the surface.
In case a particle adsorbs, it is removed from the flow and added to the respective surface element. 
In the case of recombination, an adsorbed particle is removed from the surface and the incident particle species is switched and scattered diffusively.
Every particle, which does not recombine or adsorb onto the surface, is scattered diffusively.

\subsection{Kinetic Monte Carlo Surface}
In the first step, a surface structure is replicated and the distribution of adsorbates is modelled using a Monte Carlo approach~\cite{lombardo1988monte}. 
$N_{\mathrm{x}}^2$ number of surface atoms are distributed in a structured lattice as shown in Figure \ref{fig:smcr_lattice1}.
Adsorbates are distributed on well defined sites that are defined by the number of metal atoms $n_{\mathrm{m}}$ interacting with each adsorbate.
Figure~\ref{fig:smcr_sitetypes} shows the different site types on the modelled fcc(100) surface type.
Possible site types are, on-top sites $\circled{T}$ for $n=1$, bridge sites $\circled{B}$ for $n=2$ and n-fold hollow sites $\circled{H}$ for $n>2$ surface atoms interacting with one adsorbate.
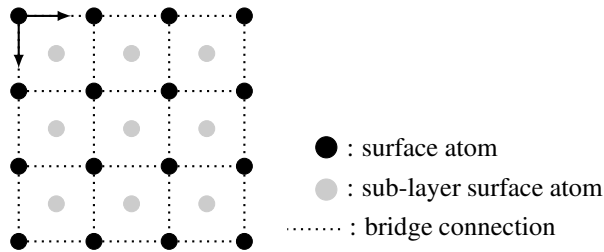
\begin{figure}[hbt!]
  \centering
  \begin{tikzpicture}
  \setcounter{owncounter}{1}
  \foreach \y in {0,...,3}
    \foreach \x in {0,...,3}
      \node[atop,scale=0.6] (\x\y) at (\x,-\y) {};% {\owncount};
  \foreach \y [count=\yi] in {0,...,2}
    \foreach \x in {0,...,3}
      \draw[thick,dotted] (\x\y)--(\x\yi) (\y\x)--(\yi\x) ;
  \foreach \y in {0,...,2}
    \foreach \x in {0,...,2}
      \node[hollowhcp,scale=0.6] at (\x+0.5,-\y-0.5) {};% {\owncount};
  %\draw[-latex] (-0.5,0.5) -- (0.5,0.5);
  %\draw[-latex] (-0.5,0.5) -- (-0.5,-0.5);
  \draw[-latex,thick] (0,0) -- (0.7,0);
  \draw[-latex,thick] (0,0) -- (0,-0.7);
\end{tikzpicture}%
  \begin{tikzpicture}
    \node[atop,scale=0.8,label=right:{: surface atom}] (topatom) at (0,0) {};
    \node[hollowhcp,scale=0.8, node distance=20pt, below of=topatom, label=right:{: sub-layer surface atom}] (sublayer) {};
    \node[hollow, node distance=15pt, below of=sublayer, label=right:{: bridge connection}] (hollow1) {};
    \node[hollow, node distance=20pt, left of=hollow1] (hollow2) {};
    \draw[thick,dotted] (hollow1.east)--(hollow2);
  \end{tikzpicture}
  \caption{On each surface element the surface is modeled by replicating its structure with one lattice of distributed atoms.
  fcc(100) type structure is modeled, exemplary shown with $N_{\mathrm{x}}=4$ surface atoms in each lattice direction.}
  \label{fig:smcr_lattice1}
\end{figure}
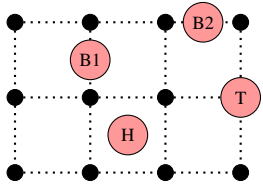
\begin{figure}[hbt!]
  \centering
  \begin{tikzpicture}
  \setcounter{owncounter}{1}
  \foreach \y in {0,...,2}
    \foreach \x in {0,...,3}
      \node[atop,scale=0.6] (\x\y) at (\x,-\y) {};% {\owncount};
  \foreach \y [count=\yi] in {0,1}
    \foreach \x in {0,...,3}
      \draw[thick,dotted] (\x\y)--(\x\yi);
  \foreach \y in {0,...,2}
    \foreach \x [count=\xi] in {0,...,2}
      \draw[thick,dotted] (\x\y)--(\xi\y);
  % put Adsorbates in correct positions
  \foreach \y in {1}
    \foreach \x in {1}
      \node[adsorbate] at (\x+0.5,-\y-0.5) {H};% {\owncount};
  \foreach \y in {1}
    \foreach \x in {3}
      \node[adsorbate] at (\x,-\y) {T};% {\owncount};
  \foreach \y in {0}
    \foreach \x in {1}
      \node[adsorbate] at (\x,-\y-0.5) {B1};% {\owncount};
  \foreach \y in {0}
    \foreach \x in {2}
      \node[adsorbate] at (\x+0.5,-\y) {B2};% {\owncount};
\end{tikzpicture}%
  \caption{Adsorbates, red filled circles, occupy sites such as on-top sites (T) for $n=1$, bridge sites (B) for $n=2$ and n-fold hollow sites (H) for $n>2$ atoms that directly interact with the
  respective adsorbate.}
  \label{fig:smcr_sitetypes}
\end{figure}

For this model we assume that all chemical processes on surfaces of the form
\begin{equation}
  \ce{ A + B ->[$r_{\ce{A-B}}$] AB} \; ,
  \label{eq:reaction_general}
\end{equation}
are described by the rate equation
\begin{equation}
  \label{eq:rate_expression_general}
  r_{\ce{A-B}} = \dfrac{d n_{\ce{A-B}}}{dt} = k_{r,\ce{A-B}} \cdot n_{\ce{A}} \cdot n_{\ce{B}} ,
\end{equation}
where $r_{\ce{A-B}}$ is the rate for the considered reaction of $\ce{A}$ with $\ce{B}$, $k_{r,\ce{A-B}}$ the rate coefficient and $n_{\ce{A}}$ and $n_{\ce{B}}$ are the respective number densities.

The reaction rate coefficient $k_{r,\ce{A-B}}$ in Equation \ref{eq:rate_expression_general} for surface reactions can be written as
\begin{equation}
  k_{r,\ce{A-B}} = \nu_r \exp \left( - \dfrac{E_{\mathrm{a}}}{k_{\mathrm{B}}T_{\mathrm{s}}} \right),
  \label{eq:rate_coeff}
\end{equation} 
with pre-exponential factor $\nu_r$, activation barrier $E_{\mathrm{a}}$, Boltzmann constant $k_{\mathrm{B}}$ and surface temperature $T_{\mathrm{s}}$.
The variables $\nu_r$ and $E_{\mathrm{a}}$ have to be defined for each reaction. % using experimental data.
Here, the missing parameters $\nu_r$ and $E_{\mathrm{a}}$ are estimated using reasonable phenomenological descriptions.

\subsubsection{activation barriers}
The activation barriers $E_{\mathrm{a}}$ in Equation \ref{eq:rate_coeff} and reaction enthalpies $\Delta h_{\mathrm{r}}$ are estimated using the UBI-QEP method~\cite{shustorovich2006ubi}: 
\begin{eqnarray}
  \Delta h_{\mathrm{r}} &=& \left[ Q_{\ce{AB}} + D_{\ce{AB}} - Q_{\ce{A}} - Q_{\ce{B}} \right] ,\\
  E_{\mathrm{a}} & = & \frac{1}{2} \left( \Delta h_{\mathrm{r}} + \dfrac{Q_{\ce{A}}Q_{\ce{B}}}{Q_{\ce{A}}+Q_{\ce{B}}} \right),
\end{eqnarray}
where $Q_{\ce{AB}}$, $Q_{\ce{A}}$ and $Q_{\ce{B}}$ are the heat of adsorption for molecular and atomic adsorbed species and $D_{\ce{AB}}$ the dissociation bond energy.
The heat of adsorption $Q_{\ce{A}}$ for simple cases such as an atomic adsorbates bound in an $n_{\mathrm{m}}$-fold site in near zero coverage limit is defined with
\begin{equation}
  Q_{\ce{A}} = Q_{\ce{0A}} \left( 2-\frac{1}{n_{\mathrm{m}}} \right)
  \label{eq:ubiqep_formalism}
\end{equation}
In order to account for lateral interactions and coverage effects, the heat of adsorption is scaled with
\begin{equation}
  \varsigma_{\mathrm{M}} = \sum_{j}^{n_{\mathrm{m}}} \frac{(2 \cdot (1/m_{j}) - (1/m_{j})^2 )}{n_{\mathrm{m}}}.
  \label{eq:scaling_local_env}
\end{equation}
Here, $m_{j}$ is the number of bound adsorbates in the nearest vicinity of the $j$th surface site-atom.
This scaling factor is used to correct the assumption of the near zero coverage limit in Equation \ref{eq:ubiqep_formalism}.

\subsubsection{pre-exponential factors}
The pre-exponential factor, which is necessary for the reaction rate coefficient (Equation \ref{eq:rate_coeff}) is estimated with the transition state 
theory~\cite{eyring1935activated,truhlar1996current} defined with
\begin{equation}
  \nu_r = \dfrac{k_{\mathrm{B}} T}{\hbar} \dfrac{\Phi_{\ce{A}^{*}}}{\Phi_{\ce{A}}},
\end{equation}
where $\hbar$ is the Planck constant and $\Phi_{\ce{A}^*}$ is the partition function for the transition state and $\Phi_{\ce{A}}$ for the adsorbed or gas state.
%Additionally the same author mentions~\cite{zhdanov1991arrhenius}: "These formulations remain under non-equilibrium conditions provided that the most favoured process pathway stays the same".
The total partition function $\Phi$ for a polyatomic species in the gas state is the product of the translational, vibrational and rotational partition functions of the considered species~\cite{cooksy2014physical}.

For the consideration of surface processes, additional assumptions are applied~\cite{zhdanov1988preexponential}.
For the partition functions of adsorbates we neglect translational and rotational states.
However, the vibrational states of the surface bonds of the adsorbates~\cite{black1982calculation} must not be neglected.
Consequently, a vibrational partition function $\Phi_{\mathrm{vib-KMCS}}$ is introduced containing two additional vibrational modes, one perpendicular and one in tangential direction to the surface.
Here, the characteristic vibrational temperatures $\Theta_{\mathrm{char-KMCS}} = Q_{\ce{A}}/(200 \cdot k_{\mathrm{B}})$ for both modes are estimated from the effective heat of adsorption $Q_{\ce{A}}$ for the considered adsorbate.

\subsubsection{Probabilities}
Probabilities for desorption processes of each adsorbate are calculated similarly to Ref.~\cite{lombardo1988monte} with
\begin{equation}
  P_{\mathrm{s}} = k_r \cdot \Delta t = \nu_r \exp \left( - \dfrac{E_{\mathrm{a}}}{k_{\mathrm{B}}T_{\mathrm{s}}} \right) \cdot \Delta t.
  \label{eq:probability_desorption}
\end{equation}
where $\Delta t$ is the chosen simulation time step.

The probability for adsorption processes is expressed as a function of the mean velocity of particles $v_{\mathrm{\perp,mean}}$ perpendicular to the surface
\begin{equation}
  \label{eq:probability_adsorption_reduced}
  %\widehat{P}_{\mathrm{ads}} = \exp \left( - \dfrac{E_{\mathrm{a}}}{k_{\mathrm{B}}T_{\mathrm{gas,ev}}} \right)
  \widehat{P}_{\mathrm{ads}} = \exp \left( - \dfrac{\pi \cdot E_{\mathrm{a}}}{2 \cdot m \cdot v_{\mathrm{\perp,mean}}^2} \right).
\end{equation}
In order to address molecular adsorption, which is a non-activated process and for which $E_{\mathrm{a,ads}}=0$ applies, we additionally consider instant desorption.
Consequently, the probability for molecular adsorption (no dissociation or recombination) yields
\begin{equation}
  \label{eq:probability_adsorption_instantdes}
  %\widehat{P}_{\mathrm{inst.-des}} = \frac{\nu_{r,ads}}{\overline{v}} \left( \exp \left( - \dfrac{E_{\mathrm{diff}}}{k_{\mathrm{B}}T_{\mathrm{gas}}} \right) \right)
  \widehat{P}_{\mathrm{molec-ads}} = 1 - \exp \left( - \dfrac{\pi \cdot E_{\mathrm{diff}}}{2 \cdot m \cdot v_{\mathrm{\perp,mean}}^2} \right),
\end{equation}
where $E_{\mathrm{diff}}$ is the diffusion barrier that we approximate with $\approx~\SI{10}{\percent}$ of the adsorption enthalpy $Q_A$

\section{The simulation tool PICLas}
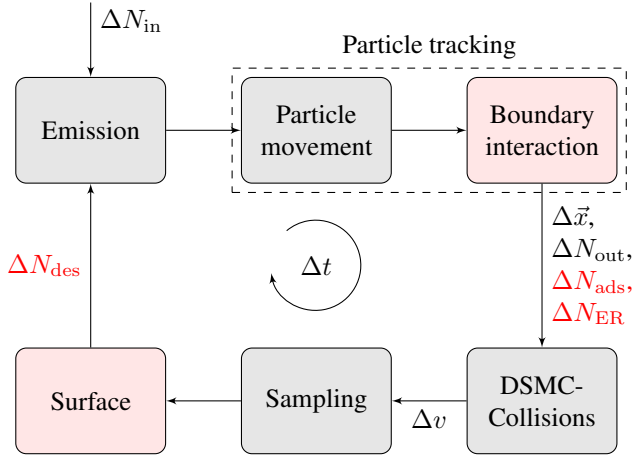
\begin{figure}[hbt!]
  \centering
  \begin{tikzpicture}[node distance = 3cm, auto]
  % Place nodes
  \node [] (center) {$\Delta t$};
  %\draw [latex'-] (0,-0.6) arc (-90:220:0.6);
  \draw [latex'-] (-0.6,-0.1) arc (-170:130:0.6);

  \node [newblock] at(3,1.8) (boundary) {Boundary interaction};
  \node [block] at(3,-1.8) (colloperator) {DSMC-Collisions};
  \node [block] at(0,-1.8) (sampling) {Sampling};
  \node [newblock] at(-3,-1.8) (surfmodel) {Surface};
  \node [block] at(-3,1.8) (emission) {Emission};
  \node [block] at(0,1.8) (move) {Particle movement};

  \node [cloudinner, fit={(move) (boundary)}, label=above:{Particle tracking}] (tracking) {};

  % Draw edges
  %\draw [line] (boundary) -- node [text width=8em] {$\Delta \vec{x}$, $\Delta N_{\mathrm{out}}$, $\Delta N_{\mathrm{ads}}$, $\Delta N_{\mathrm{ER}}$} (surfupdate);
  %\path [line] (surfupdate) -- (colloperator);
  \draw [line] (boundary) -- node [text width=4em] {$\Delta \vec{x}$, $\Delta N_{\mathrm{out}}$, \color{red}{$\Delta N_{\mathrm{ads}}$, $\Delta N_{\mathrm{ER}}$}} (colloperator);
  \path [line] (colloperator) -- node {$\Delta v$} (sampling);
  \path [line] (sampling) -- (surfmodel);
  \path [line] (surfmodel) -- node {\color{red}{$\Delta N_{\mathrm{des}}$}} (emission);
  \path [line] (emission) -- (move);
  \path [line] (move) -- (boundary);

  \path [line] (-3,3.5) -- node [near start] {$\Delta N_{\mathrm{in}}$} (emission);
\end{tikzpicture}
  \caption{Scheme of a time step loop in PICLas. Collisions are solved with the DSMC module. Surface models add-ons are depicted red.}
  \label{fig:flowchart_piclas}
\end{figure}

We use the particle simulation tool PICLas~\cite{pic_dsmc_ueberblick} for simulations.
Figure~\ref{fig:flowchart_piclas} shows how a representative time step is realized.
First the particles are inserted into the domain.
Then, each time step starts with an emission step where particles are inserted at an inflow surfaces using a surface flux~\cite{Garcia2006GenerationOT}.
After the emission, particles are moved within the simulation domain and their positions are updated.
This is combined with a boundary interaction routine in the particle tracking block.
During these interactions, the location of particles intersecting boundaries is updated or the particles are removed from the domain.
Following, inter particle collisions are treated with the DSMC method, where velocity and internal energies of the particles are changed.
In the last step the macroscopic values are sampled and an output is generated.
The changes and add-ons performed in the project context are highlighted red and are described in the following.

\section{Simulations}
\label{sec:simulations}
% ============================================================================
% Plasma wind tunnel flow ====================================================
% ============================================================================
\subsection{Setup}
\begin{figure}[hbt!]
  \centering
  \begin{tikzpicture}
  \begin{axis}[ width=0.49\textwidth,
                height=0.28\textheight,
                view={0}{0},
                grid=both,
                ymin=-80,
                ymax=2080,
                xmin=-80,
                xmax=2150,
                xlabel={$x \; (\si{\centi\metre})$},
                ylabel={$r \; (\si{\centi\metre})$},
                xtick={0,400,800,1200,1600,2000},
                xticklabels={-8,-6,-4,-2,0,2},
                ytick={0,500,1000,1500,2000},
                yticklabels={0,2,4,6,8},
                tick label style ={font=\footnotesize},
                colorbar style={font=\footnotesize},
              ]
    \addplot graphics [xmin=0,xmax=2100,ymin=0,ymax=2000] {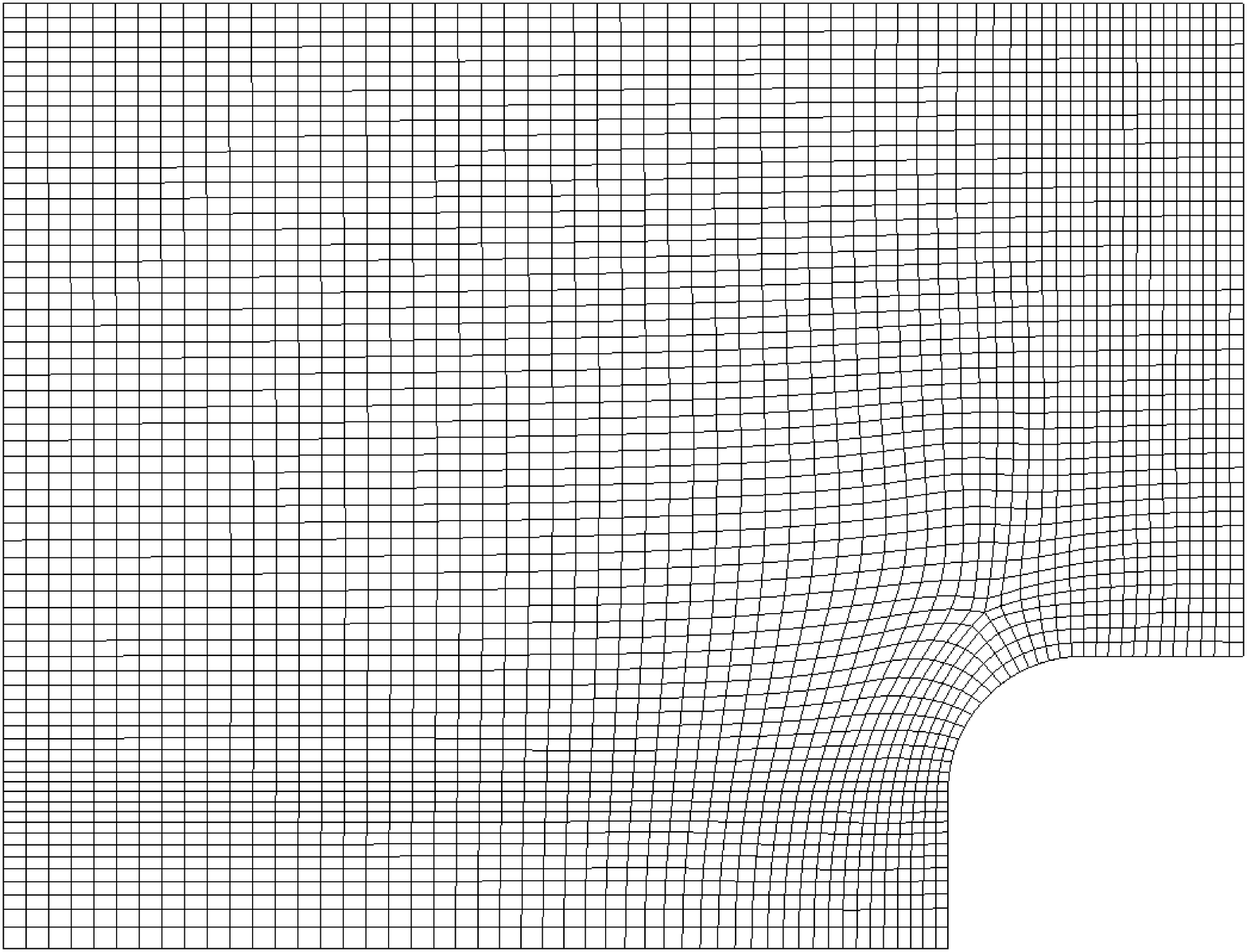};
  \end{axis}
\end{tikzpicture}
  \caption{A radial slice of the mesh.}
  \label{fig:simulation_setup1_slice}
\end{figure}

\begin{figure}[hbt!]
  \centering
  \begin{tikzpicture}
  \begin{axis}[ width=0.45\textwidth,
                view={0}{0},
                hide axis,
                clip = false,
                ymin=0,
                ymax=2075,
                xmin=0,
                xmax=2800
              ]
    \addplot graphics [xmin=0,xmax=2800,ymin=0,ymax=2075] {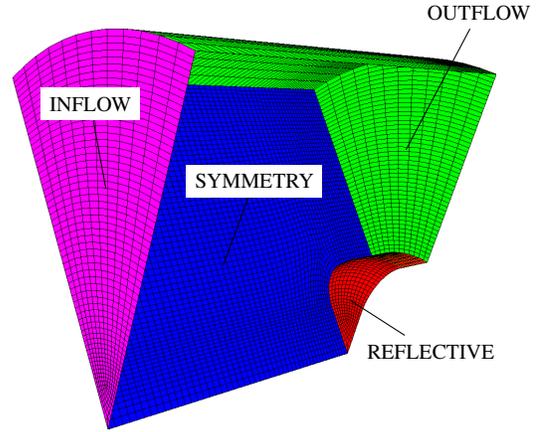};
    \node[fill=white, anchor=south, style={font=\footnotesize}] (source1) at (axis cs:2700,2075){OUTFLOW};
    \node (destination1) at (axis cs:2250,1400){};
    \draw[-](source1)--(destination1);
    \node[fill=white, anchor=south, style={font=\footnotesize}] (source2) at (axis cs:450,1600){INFLOW};
    \node (destination2) at (axis cs:550,1200){};
    \draw[-](source2)--(destination2);
    \node[fill=white, anchor=south, style={font=\footnotesize}] (source3) at (axis cs:1400,1200){SYMMETRY};
    \node (destination3) at (axis cs:1200,800){};
    \draw[-](source3)--(destination3);
    \node[fill=white, anchor=west, style={font=\footnotesize}] (source4) at (axis cs:2000,400){REFLECTIVE};
    \node (destination4) at (axis cs:1900,700){};
    \draw[-](source4)--(destination4);

  \end{axis}
\end{tikzpicture}
  \caption{The 3D simulation domain with defined boundaries.}
  \label{fig:simulation_setup1}
\end{figure}

In the next step, PICLas simulations are presented based on a setup from plasma wind tunnel (PWT) experiments~\cite{massuti_jtht2017}.
For this, a mesh is created for the ESA standard pressure probe.
PICLas uses 3D hexahedral meshes, and here it is build using GridPro~\cite{gridpro_manual}.
The setup consist of a $\approx\,\SI{40}{\degree}$ cylindrical domain to reduce computational effort.
Figures \ref{fig:simulation_setup1_slice} and \ref{fig:simulation_setup1} show a representative mesh slice as well as the boundary conditions.

Three simulations are performed using first non-catalytic then fully catalytic surfaces and last kinetic Monte Carlo surfaces.
The inflow conditions are calculated from an experimental flow characterization~\cite{massuti_jtht2017} and are given in Table~\ref{tab:inflow_parameters}.
Rotational and vibrational temperatures are assumed to be equal to translational temperature.
An adaptive pressure boundary~\cite{farbar2014subsonic} with $p_{\infty}=p_{\infty,\mathrm{inflow}}$ is applied to the outflow condition because of low Mach numbers.
Dissociation bond energy of oxygen $D_{\ce{O2}} = \SI{498.34}{\kilo\joule\per\mol}$ is taken from Ref.~\cite{crc}, the Arrhenius coefficients for oxygen gas reactions from Ref.~\cite{park1994review}, and
the remaining physical parameters such as characteristic vibrational temperatures are taken from the NIST database.
For the VHS collision model in the DSMC method the chosen parameters are the VHS exponent $\omega_{\mathrm{VHS}}=\num{0.77}$~\cite{bird1994molecular} and reference temperature
$T_{\mathrm{ref}}=\SI{273}{\kelvin}$ resulting in the reference diameters $d_{\mathrm{ref,\ce{O2}}}=\SI{4.07e-10}{\metre}$ and $d_{\mathrm{ref,\ce{O}}}=\SI{3.11e-10}{\metre}$.
The particle weighting factor and simulation time step are chosen with $W_{\mathrm{p}}=\num{1e10}$ and $\Delta t=\SI{5e-8}{\second}$.
\begin{table}[hbt!]
  \centering
  \caption{Parameters calculated for inflow boundary at the center of inflow ($r = \SI{0}{\metre}$). Parameters are static pressure, total pressure, translational temperature, surface temperature,
  particle number density for $\ce{O}$ and $\ce{O2}$, velocity in x-direction, Mach number at inflow and global Knudsen number in that order.}
  \tabcolsep7pt\vspace{2mm}
  \begin{tabular}{lll}
    \hline
    \hline
    Property                   & Unit                     & Value \\
    \hline
    $p_{\infty}$               & \si{\pascal}           & \num{40} \\
    $p_{\mathrm{tot}}$         & \si{\pascal}           & \num{149} \\
    $T_{\mathrm{tr}}$          & \si{\kelvin}           & \num{3342} \\
    $T_{\mathrm{s},\ce{SiO2}}$ & \si{\kelvin}           & \num{1398} \\
    $n_{\ce{O}}$               & \si{\per\cubic\metre}  & \num{8.15E+20} \\
    $n_{\ce{O2}}$              & \si{\per\cubic\metre}  & \num{5.18E+19} \\
    $v_{x}$                    & \si{\metre\per\second} & \num{2412.3} \\
    $Ma_{\infty}$              & -                      & $\approx$ \num{1.44} \\
    $Kn_{\mathrm{VHS}}$        & -                      & $\approx$ \num{0.01} \\
    \hline
    \hline
  \end{tabular}
  %\begin{tabular}{lllllllll}
    %\hline
    %Inflow Set & $p_{\infty}$ [\SI{}{\pascal}]  & $p_{\mathrm{tot}}$ [\SI{}{\pascal}] & $T_{\mathrm{tr}}$ [\SI{}{\kelvin}] & $n_{\mathrm{O}}$ [\SI{}{\per\cubic\metre}]&  $n_{\mathrm{O}_2}$ [\SI{}{\per\cubic\metre}] & $v_{x}$ [\SI{}{\metre\per\second}] & $Ma_{\infty}$ [-]  & $Kn_{\mathrm{VHS}}$ [-]\\
    %\hline
    %1 & \num{40} & \num{120} & \num{3985} & \num{6.19E+20} & \num{1.07E+20} & \num{2218.0} & $\approx$ \num{1.30} & $\approx$ \num{0.01} \\
    %2 & \num{40} & \num{149} & \num{3342} & \num{8.15E+20} & \num{5.18E+19} & \num{2412.3} & $\approx$ \num{1.44} & $\approx$ \num{0.01} \\
    %3 & \num{160} & \num{460} & \num{4985} & \num{1.29E+21} & \num{1.04E+21} & \num{2130.0} & $\approx$ \num{1.29} & $\approx$ \num{0.003} \\
    %\hline
  %\end{tabular}
  \label{tab:inflow_parameters}
\end{table}

The flow inside the simulation domain is initialized with two species (\ce{O}, \ce{O2}), uniformly distributed using the inflow parameters.
The total simulation duration is chosen with $t_{\mathrm{end}}=\SI{1.4e-3}{\second}$.
At $t_{sim}=\SI{1e-3}{\second}$ the flow and surface values have reached steady state after which sampling is enabled.
From this moment the sampling is enabled for $t_{\mathrm{sample}}=\SI{4e-4}{\second}$.

\begin{figure}[hbt!]
  \centering
  \begin{tikzpicture}
  \foreach \x in {0,...,3}
    \foreach \y in {0,...,2}
      \node[atop,scale=0.8] (\x\y) at (\x,-\y) {};
  \foreach \x in {0,...,3}
    \foreach \y [count=\yi] in {0,1}
      \draw[thick,dotted] (\x\y)--(\x\yi);
  \foreach \x [count=\xi] in {0,...,2}
    \foreach \y in {0,...,2}
      \draw[thick,dotted] (\x\y)--(\xi\y);
  % put Oxygen atoms in correct positions
  \foreach \x in {0,...,3}
    \foreach \y in {0,1}
      \node[adsorbate2,scale=0.6] at (\x,-\y-0.5) {};
  \foreach \x in {0,...,2}
    \foreach \y in {0,...,2}
      \node[adsorbate2,scale=0.6] at (\x+0.5,-\y) {};
\end{tikzpicture}%
  \begin{tikzpicture}
    \node[atop,scale=0.8,label=right:{: \ce{Si}-atom}] (topatom) at (0,0) {};
    \node[adsorbate2,scale=0.8, node distance=20pt, below of=topatom, label=right:{: \ce{O}-atom}] (sublayer) {};
    \node[hollow, node distance=15pt, below of=sublayer, label=right:{: bridge connection}] (hollow1) {};
    \node[hollow, node distance=20pt, left of=hollow1] (hollow2) {};
    \draw[thick,dotted] (hollow1.east)--(hollow2);
  \end{tikzpicture}
  \caption{The pre-oxidized \ce{SiC} surface (\ce{SiO2}) is represented by a reconstructed silicate substrate. \ce{Si} surface atoms form a fcc(100) structure with \ce{O} bound in 2-fold (bridge) sites.}
  \label{fig:sio2_lattice}
\end{figure}
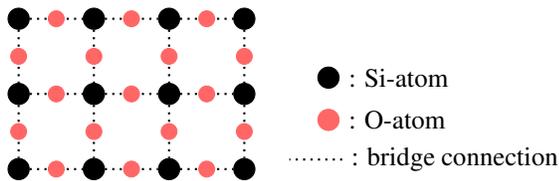

Although the UBI-QEP model for the estimation of activation barriers is constructed for transitional metals~\cite{shustorovich2007ubiqep_vs_dft}, we use the same formalism here.
The pre-oxidized \ce{SiC}.
Recombination reactions of gas phase oxygen ($\ce{O}^{(\mathrm{gas})}$) on \ce{SiO2} surfaces are considered to be reactions of gas phase $\ce{O}^{(\mathrm{gas})}$
with bound $\ce{O}^{(\mathrm{surf})}$ in the \ce{SiO2} lattice~\cite{seward1991model}.
Consequently, the catalytic surface is taken to be a reconstructed silicate substrate rather than \ce{SiO2} and we replicate it with a \ce{Si} surface lattice
with \ce{Si}-atom distances present in \ce{SiO2} ($\sigma \approx \SI{2.6156e+18}{\per\square\metre}$).
Derived from Ref.~\cite{schwiete1958gitterenergie} bond orders of \ce{Si} and \ce{O} in the \ce{SiO2} crystal are taken to be $n_{m,\ce{O}}=2$ and $m_{j,\ce{Si}}=4$.
This forms a fcc(100) type surface structure as shown in figure~\ref{fig:sio2_lattice} with \ce{O} bound in 2-fold (bridge) sites.
The discretisation on each surface element is realised with $N_\mathrm{x}=40$.
The dissociation bond energy $D_{\ce{Si-O}} = \SI{798}{\kilo\joule\per\mol}$~\cite{crc} of the \ce{Si-O} bond is used as heat of adsorption $Q_{\ce{0A}}$ for atomic oxygen.
Applying equations \ref{eq:ubiqep_formalism} and \ref{eq:scaling_local_env}, we derive $Q_{\ce{A}}$ on the reconstructed \ce{Si} that represents a \ce{SiO2} surface
for an oxygen surface coverage of $\theta_{\ce{O}/\ce{Si}}=2$.
This derived $Q_{\ce{A}}=\SI{451.81}{\kilo\joule\per\mol}$ also represents the value given by Ref.~\cite{wells2012structural} for the dissociation bond energy of the \ce{O-SiO} bond in the real
\ce{SiO2} surface.

\subsection{Results}
First, we would like to comment on the comparison of simulation results and experimental data.
The simulation result is strongly dependant on the defined inflow condition for such a plasma flow.
The experimental flow characterization~\cite{massuti_jtht2017} with which these simulations are compared has many dependencies and is subject to a combination of uncertainties.
Consequently, this can lead to great deviations of the simulation results and experimental measurements at the probe surface.

\subsubsection{Pressure on probe surface}
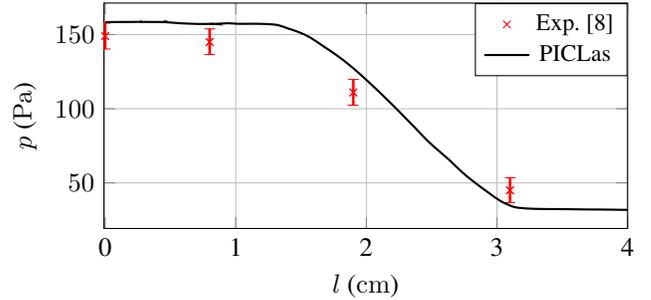
\begin{figure}[hbt!]
  \centering
  %plotting surface pressure
\begin{tikzpicture}%
  \begin{axis}[ legend pos=north east,
                legend style = {at={(1,1)},font=\small},
                width=0.48\textwidth,
                height=0.2\textheight,
                grid=both,
                xmin=-1e-4,
                xmax=0.04,
                xlabel={$l$ ($\si{\centi\metre}$)},
                ylabel={$p$ ($\si{\pascal}$)},
                xtick={0,1e-2,2e-2,3e-2,4e-2},
                xticklabels={0,1,2,3,4},
                scaled ticks=false,
]
    %\addplot [only marks, mark=x,    color=red] coordinates {(0,142) (0.01,136) (0.019,107) (0.031,43)};
    \addplot[forget plot, only marks, mark=x,    color=red, error bars/.cd, y dir=plus, y explicit] table [col sep=comma, x=x, y=pressure, y error expr={\thisrow{max}-\thisrow{pressure}}] {tikz/PWK_Setups/Setup1/Data/exp_surf_pressure.csv};
    \addplot[only marks, mark=x,    color=red, error bars/.cd, y dir=minus, y explicit] table [col sep=comma, x=x, y=pressure, y error expr={\thisrow{pressure}-\thisrow{min}}] {tikz/PWK_Setups/Setup1/Data/exp_surf_pressure.csv};
    \addlegendentry{Exp.~\cite{massuti_jtht2017}}
    \addplot [mark=none, color=black, solid, thick, smooth] table [col sep=comma, x=arc_length, y expr=sqrt((\thisrow{ForcePerArea_0})^2+(\thisrow{ForcePerArea_1})^2+(\thisrow{ForcePerArea_2})^2)] {tikz/PWK_Setups/Setup1/Data/surfaceData_M1.csv};
    \addlegendentry{PICLas}
  \end{axis}
\end{tikzpicture}%
  \caption{Pressure distribution along the probe surface for the chosen inflow conditions.}
  \label{fig:pressure_pwt}
\end{figure}

The simulated pressure along the probe surface is shown in figure \ref{fig:pressure_pwt} and compared to experimental measurements.
Simulated values for the pressure are above the experimental values but predict a similar progress along the surface.

\subsubsection{Convective reactive heat fluxes on the probe surface}
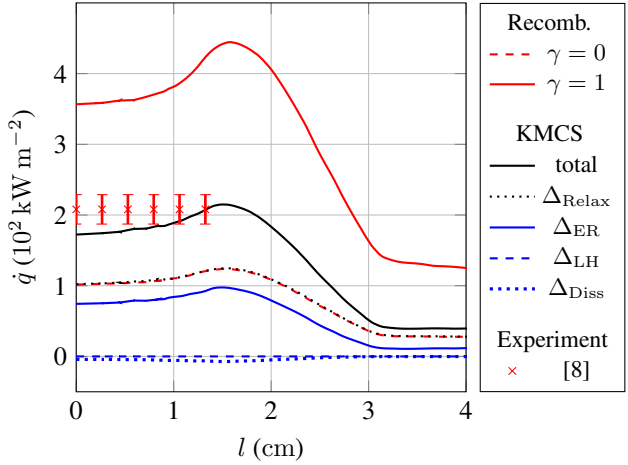
\begin{figure}[hbt!]
  \centering
  %plotting heat flux
\begin{tikzpicture}
  \begin{axis}[ legend pos=north west,
                legend style = {at={(1.035,1)},font=\small},
                width=0.38\textwidth,
                height=0.295\textheight,
                grid=both,
                ymin=-5e4,
                ymax=5e5,
                xmin=0,
                xmax=0.04,
                xlabel={$l$ ($\si{\centi\metre}$)},
                ylabel={$\dot{q}$ ($\SI{e2}{\kilo\watt\per\square\metre}$)},
                ytick={0,1e5,2e5,3e5,4e5},
                yticklabels={0,1,2,3,4},
                xtick={0,1e-2,2e-2,3e-2,4e-2},
                xticklabels={0,1,2,3,4},
                scaled ticks=false,
]
    \addlegendimage{empty legend}\addlegendentry{\hspace{-2.5em} Recomb.}
    \addplot [mark=none, color=red, dashed, thick, smooth] table [col sep=comma, x=arc_length, y=HeatFlux] {tikz/PWK_Setups/Setup1/Data/surfaceData_M1.csv};
    \addlegendentry{$\gamma=0$}
    \addplot [mark=none, color=red, solid, thick, smooth] table [col sep=comma, x=arc_length, y=HeatFlux] {tikz/PWK_Setups/Setup1/Data/surfaceData_M2fc.csv};
    \addlegendentry{$\gamma=1$}
    \addlegendimage{empty legend}\addlegendentry{\color{white}|}
    \addlegendimage{empty legend} \addlegendentry{\hspace{-2.5em} KMCS}
    \addplot [mark=none, color=black, solid, thick, smooth] table [col sep=comma, x=arc_length, y=HeatFlux] {tikz/PWK_Setups/Setup1/Data/surfaceData_M3.csv};
    \addlegendentry{total}
    \addplot [mark=none, color=black, dotted, thick, smooth] table [col sep=comma, x=arc_length, y expr=(\thisrow{HeatFlux}-\thisrow{HeatFlux_Portion_ER}-\thisrow{HeatFlux_Portion_LH}-(\thisrow{HeatFlux_Portion_SurfDiss}+\thisrow{HeatFlux_Portion_AdsDiss}))] {tikz/PWK_Setups/Setup1/Data/surfaceData_M3.csv};
    \addlegendentry{$\Delta_{\mathrm{Relax}}$}
    \addplot [mark=none, color=blue, solid, thick, smooth] table [col sep=comma, x=arc_length, y=HeatFlux_Portion_ER] {tikz/PWK_Setups/Setup1/Data/surfaceData_M3.csv};
    \addlegendentry{$\Delta_{\mathrm{ER}}$}
    \addplot [mark=none, color=blue, dashed, thick, smooth] table [col sep=comma, x=arc_length, y=HeatFlux_Portion_LH] {tikz/PWK_Setups/Setup1/Data/surfaceData_M3.csv};
    \addlegendentry{$\Delta_{\mathrm{LH}}$}
    \addplot [mark=none, color=blue, dotted, very thick, smooth] table [col sep=comma, x=arc_length, y expr=(\thisrow{HeatFlux_Portion_SurfDiss}+\thisrow{HeatFlux_Portion_AdsDiss})]{tikz/PWK_Setups/Setup1/Data/surfaceData_M3.csv};
    \addlegendentry{$\Delta_{\mathrm{Diss}}$}
    \addlegendimage{empty legend}\addlegendentry{\color{white}|}
    \addlegendimage{empty legend}\addlegendentry{\hspace{-2em}Experiment}
    \addplot [only marks, mark=x, color=red, solid, error bars/.cd, y dir=both, y fixed relative=0.1] coordinates {(0,208000) (0.00265,208000) (0.00530,208000) (0.00795,208000) (0.01060,208000) (0.01325,208000)};
    \addlegendentry{\cite{massuti_jtht2017}}
  \end{axis}
\end{tikzpicture}
  \caption{Heat fluxes onto the probe surface.
  First simulation results achieved with the simple recombination model are depicted. Secondly, results using the KMCS method split into separate reaction portions.
  Third, experiment measurements.}
  \label{fig:heatfluxes_set2}
\end{figure}

The heat fluxes along the surface for all simulations are shown in figure~\ref{fig:heatfluxes_set2}.
The results are separated into three parts.
First, the simulation results for non-catalytic $\gamma=0$ and fully catalytic surfaces $\gamma=1$ are listed in red.
Here, the minimum and maximum heat fluxes for this flow are achieved.
Secondly, the simulation result using the KMCS method are presented.
These results are split into four portions, one for each reaction type.
$\Delta_{\mathrm{Relax}}$ shows the portion resulting from relaxation of colliding particles at the surface.
This heat flux portion is equal to the non-catalytic ($\gamma=0$) case.
$\Delta_{\mathrm{ER}}$ combines all heat fluxes resulting from recombination reactions of incident particles and adsorbates.
$\Delta_{\mathrm{LH}}$ combines all heat fluxes resulting from recombination reactions of two adsorbates with subsequent desorption.
$\Delta_{\mathrm{Diss}}$ combines all heat fluxes resulting from dissociation reactions on the surface.
The sum of all these heat fluxes yields the total heat flux resulting from using the KMCS method.
The third part added to figure~\ref{fig:heatfluxes_set2} is the experimentally measured data value that represents the integral heat-flux onto the forward facing surface.
It is illustrated with six data points, each one at $\dot q_{\mathrm{exp}}~=~\SI{208}{\kilo\watt\per\square\metre}$.

With the KMCS method the heat flux $\dot q_{\mathrm{KS-total}}$ is estimated between the heat fluxes for $\gamma=0$ and $\gamma=1$.
%The heat flux simulated with the KMCS method $\dot q_{\mathrm{KS-total}}$ is positioned in between the heat fluxes for $\gamma=0$ and $\gamma=1$.
It is in a good agreement with experimentally measured values, considering the uncertainties of the chosen inflow condition.
The heat flux portions show that most of the reactive heat flux results from an ER-reaction of oxygen atoms on the surface.
A small part of the colliding \ce{O2} molecules dissociate on the \ce{SiO2} surface, which slightly decreases the heat flux.

Here, the effective recombination coefficient is calculated with
\begin{equation}
  \gamma_{\mathrm{eff}} = \dfrac{2 \cdot N_{\ce{O}\mathrm{,LH}} + 2 \cdot N_{\ce{O}\mathrm{,ER}} }{2 \cdot N_{\ce{O2}\mathrm{,Diss}} } \cdot \dfrac{1}{N_{\ce{O}\mathrm{,Coll}}},
  \label{eq:gamma_eff}
\end{equation}
where $N_{\ce{O}\mathrm{,LH}}$, $N_{\ce{O}\mathrm{,ER}}$, and $N_{\ce{O2}\mathrm{,Diss}}$ are the number of the respective reactions and $N_{\ce{O}\mathrm{,Coll}}$ the number of surface collisions.
%\begin{figure}[t!]
\begin{figure}[hbt!]
  \centering
  %plotting surface recombination coefficient
\begin{tikzpicture}
  \begin{axis}[ legend pos=north east,
                legend style = {at={(1,1)},font=\small},
                width=0.49\textwidth,
                height=0.2\textheight,
                %legend columns=2,
                %legend style={
                  % the /tikz/ prefix is necessary here...
                  % otherwise, it might end-up with `/pgfplots/column 2`
                  % which is not what we want. compare pgfmanual.pdf
                    %/tikz/column 2/.style={column sep=10pt,},
                  %},
                grid=both,
                xmin=0,
                xmax=0.04,
                xlabel={$l$ ($\si{\centi\metre}$)},
                ylabel={$\gamma_{\mathrm{eff}}$ ($\si{-}$)},
                xtick={0,1e-2,2e-2,3e-2,4e-2},
                xticklabels={0,1,2,3,4},
                scaled ticks=false,
                scaled y ticks={base 10:1},
               ]
    \addplot [forget plot, mark=none, color=black, thick, smooth] table
    [col sep=comma, x=arc_length, y expr=((2*\thisrow{Spec002_CollReact002_Count}+2*\thisrow{Spec002_SurfReact002_Count})/(\thisrow{Spec002_Counter}+
    2*\thisrow{Spec001_CollReact001_Count}+2*\thisrow{Spec001_SurfReact001_Count}))] {tikz/PWK_Setups/Setup1/Data/surfaceData_M3.csv};
    \addplot [forget plot, mark=none, color=red, thick, smooth, domain=0:0.04]{0.11};
    \addlegendimage{mark=none, color=red, thick}\addlegendentry{Massuti~\cite{massuti_jtht2017}}
    \addlegendimage{mark=none, color=black, thick}\addlegendentry{PICLas}
  \end{axis}
\end{tikzpicture}%
  \caption{Recombination coefficient along the probe surface.}
  \label{fig:gamma}
\end{figure}
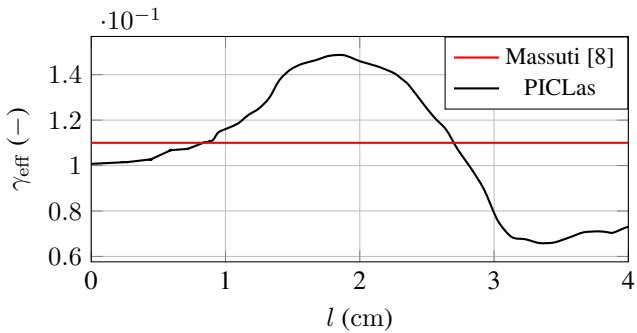
Figure~\ref{fig:gamma} shows the effective recombination coefficient that results from the PICLas simulation.
In the region of interest ($l=\num{0}-\SI{4}{\centi\metre}$) the recombination coefficient takes the values \num{0.07} to \num{0.15}.
This is in a good agreement with the reference value of \num{0.11} for this setup.

\subsubsection{Computational times}
For the shown simulations we analyzed the computational times.
Each computation was performed on 200 Intel Xeon E5-2670 v2 (Sandy Bridge) processors.
Table \ref{tab:compute_times} shows the computational times and increase in computational time using catalytic models.
These results show almost no increase in computational cost while using the recombination model.
This is expected, since this model only adds one additional check prior to reflection of the recombining atoms.
In contrast, using the KMCS method increases the computational cost by \SI{40}{\percent} in comparison to simulations without enabled catalysis modelling.
This results from a profound modelling and discretisation on every surface.
\begin{table}[hbt!]
  \centering
  \caption{Computational times given in CPUh for simulations with different surface models applied.}
  \tabcolsep6pt\vspace{2mm}
  \begin{tabular}{ccc|c}
    \hline
      DSMC & Recomb & KMCS & scaling \\
      ${C1}$ & ${C2}$ & ${C3}$ & $(C3-C1)/C1$\\
    \hline
      305 & 309 & 423 & 0.39 \\
    \hline
  \end{tabular}
  \label{tab:compute_times}
\end{table}

\section{Conclusion}
Statistical particle methods such as the DSMC method are well established for design purposes of TPS.
In order to predict catalytic flows these methods use reactive surface models.

Two modelling approaches that treat surface reactions were implemented into the plasma flow solver PICLas.
The first model accounts for simple recombination of atomic species on the surface.
The second model is a kinetic Monte Carlo approach where a surface structure on each surface element is replicated and reactions on the replicated surface are modelled statistically.
This model is intended to be independent of parameter fitting and reaction parameters are estimated with phenomenological assumptions.
In contrast to the simple recombination model a broad range of reactions including Eley-Rideal, Langmuir-Hinshelwood and dissociation reactions can be defined.

The ability of each model to predict the flow and surface reactivity was tested by conducting a simulation of a catalytic oxygen flow over an ESA standard pressure probe.

The recombination model leads to only a slight increase in the computational effort.
However, the application of the recombination model is limited to flows where the recombination coefficient was fitted beforehand,
or for simulations with the goal to estimate the maximum and minimum bounds of the heat loads.

The KMCS method is capable of estimating the heat loads that result from surface reactions with a good agreement to experiments without priorly fitting of parameters.
This also applies to the effective recombination coefficient.
A challenge of the KMCS method is the need to phenomenologically replicate the occurring processes at the surface before conducting simulations.
This includes surface structure, adsorbate binding-sites and binding-energies.
In this work, the chosen structural surface parameters for oxygen flows over \ce{SiO2} surfaces are well suited to reproduce the considered plasma flow.
The computational effort increases by \SI{40}{\percent} with the chosen discretisation.

%\bibliographystyle{IEEEbib}
%\bibliography{refs}

\end{document}